\documentclass[10pt,twocolumn,letterpaper]{article}

\usepackage{cvpr}
\usepackage{times}
\usepackage{epsfig}
\usepackage{graphicx}
\usepackage{amsmath}
\usepackage{amssymb}
\usepackage{booktabs}
\usepackage{makecell}
\usepackage{hyperref}   % For hyperlinks and references
\usepackage{graphicx}   % For images
\usepackage{subcaption}
\usepackage{algorithm}
\usepackage[noend]{algorithmic}
\usepackage{paralist}

\begin{document}
\setcounter{footnote}{0}
\renewcommand{\thefootnote}{\fnsymbol{footnote}}

%%%%%%%%% TITLE
\title{Reconstructing Protected Biometric Templates from Binary Authentication Results}

\author{
Eliron Rahimi\textsuperscript{1} \quad 
Margarita Osadchy\textsuperscript{2} \quad 
Orr Dunkelman\textsuperscript{2} \\
Computer Science Department, University of Haifa, Haifa, Israel \\
\textsuperscript{1}{\tt\small elironrahimiacademy@gmail.com} \quad
\textsuperscript{2}{\tt\small \{rita, orrd\}@cs.haifa.ac.il}
}

\maketitle

\begingroup
\renewcommand{\thefootnote}{}
\footnotetext{Accepted at the International Joint Conference on Biometrics (IJCB) 2025.}
\endgroup

\thispagestyle{empty}

%%%%%%%%% ABSTRACT
\begin{abstract}

Biometric data is considered to be very private and highly sensitive. As such, many methods for biometric template protection were considered over the years --- from biohashing and specially crafted feature extraction procedures, to the use of cryptographic solutions such as Fuzzy Commitments or the use of Fully Homomorphic Encryption (FHE).

A key question that arises is how much protection these solutions can offer when the adversary can inject samples, and observe the outputs of the system. While for systems that return the similarity score, one can use attacks such as hill-climbing, for systems where the adversary can only learn whether the authentication attempt was successful, this question remained open.

In this paper, we show that it is indeed possible to reconstruct the biometric template by just observing the success/failure of the authentication attempt (given the ability to inject a sufficient amount of templates). Our attack achieves negligible template reconstruction loss and enables full recovery of facial images through a generative inversion method, forming a pipeline from binary scores to high-resolution facial images that successfully pass the system more than 98\% of the time. Our results, of course, are applicable for any protection mechanism that maintains the accuracy of the recognition.

\end{abstract}

%%%%%%%%% BODY TEXT

% \begingroup
% \renewcommand{\thefootnote}{}
% \footnotetext{Accepted at the International Joint Conference on Biometrics (IJCB) 2025.}
% \endgroup

\section{Introduction}
\label{sec:Introduction}

Thanks to advances in deep learning, biometric authentication systems are widely utilized across various application scenarios and are expected to see even broader adoption over time. Although biometric authentication provides high usability and security due to the descriptiveness and
uniqueness of biometric features, it also presents challenges related to privacy protection for enrolled subjects~\cite{nita2018security}.

It has been shown that face images can be reconstructed from templates generated by deep feature extraction algorithms~\cite{dong2023reconstruct,gomez2012face,mai2018reconstruction,vendrow2021realistic}, highlighting the potential vulnerability of biometric templates. The European Union’s General Data
Protection Regulation (GDPR)~\cite{regulation2016regulation} declared
biometric data as sensitive personal
data and
\begin{figure}[ht]
    \centering
    \includegraphics[width=1\linewidth]{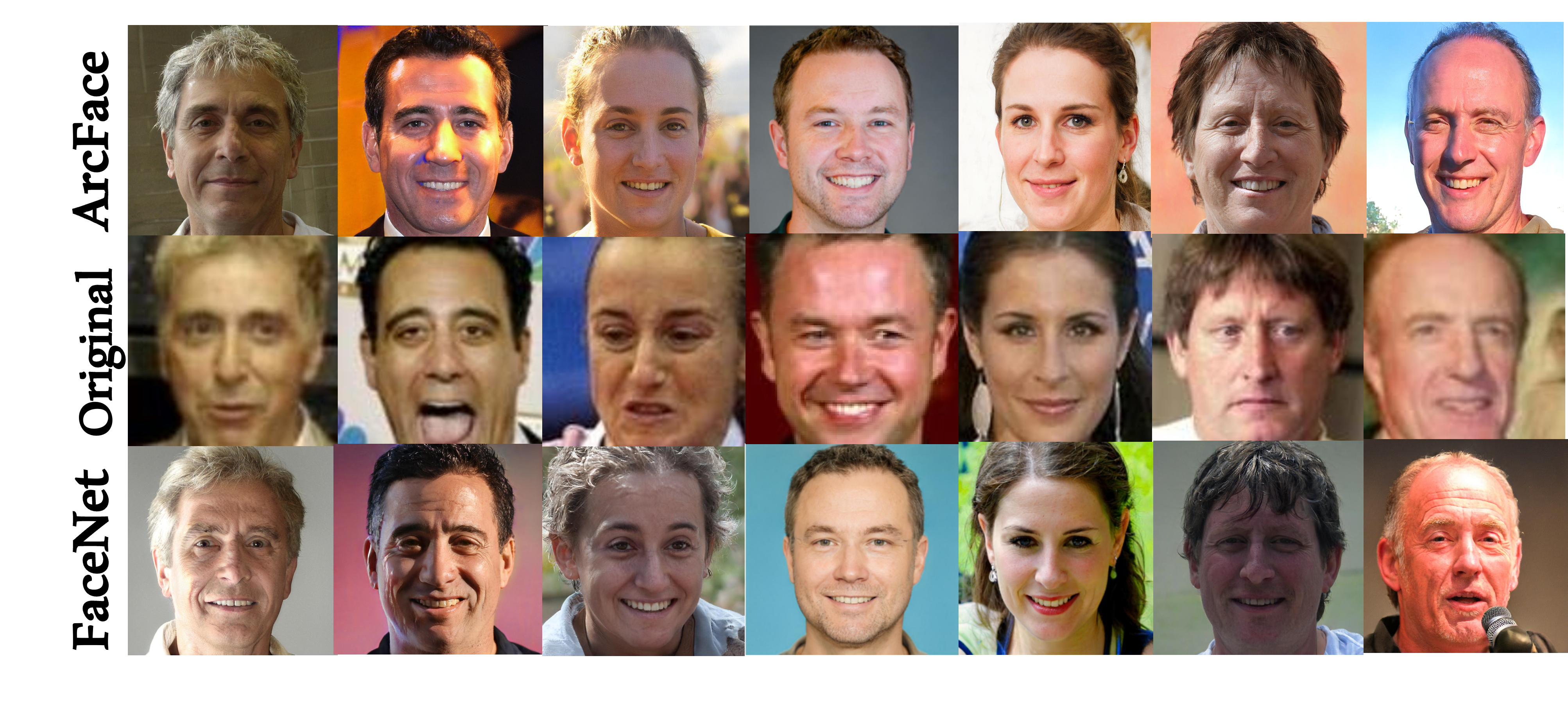}
    \caption{Reconstructed images obtained by our attack from protected ArcFace and FaceNet templates compared to the original images.}
    \label{fig:reconstruct}
\end{figure}
ISO/IEC 24745~\cite{iso2011iec} provided a standardized framework for protecting biometric information through the requirements of unlinkability, irreversibility and renewability. In this work we address the irreversibility requirement.  
Many different protection mechanisms including well known cryptographic solutions such as Fully Homomorphic Encryption (FHE), Secure Multiparty Computation (MPC), and Attribute-Based Encryption (ABE) have been suggested for biometric templates~\cite{HahnM23}. FHE~\cite{yi2014homomorphic} has been considered one of the most promising among them due to its ability to perform arithmetic operations such as addition and multiplication in the encrypted domain. Hence, one can compare the new template with the encrypted template, without decrypting the template and exposing the original biometric data with minimal accuracy loss~\cite{bassit2021fast,bassit2023improved,bauspiess2022improved,boddeti2018secure,engelsma2022hers,kolberg2020efficiency,morampudi2021secure,pradel2021privacy,tamiya2021improved,yang2020secure}.

\begin{figure}[ht]
    \centering
    
    % Subfigure (a) - Enrollment process
    \begin{subfigure}[b]{0.5\textwidth}
        
        \includegraphics[width=\textwidth]{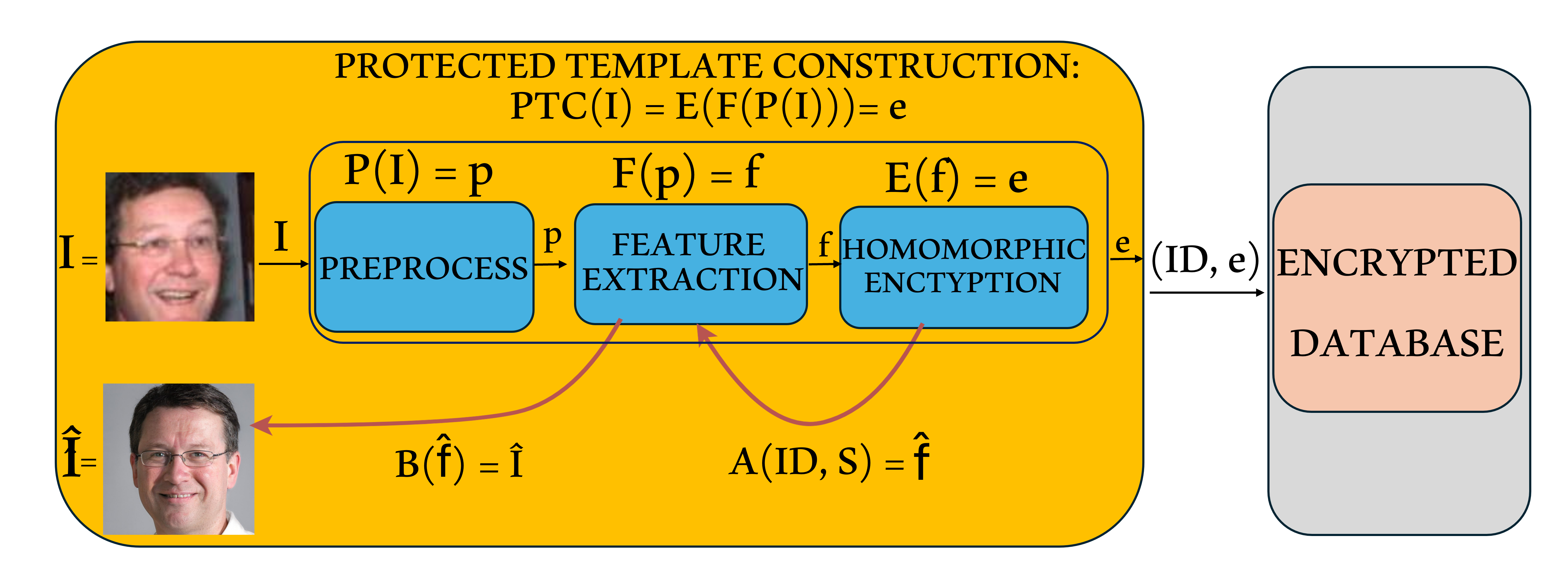}
        \caption{Standard enrollment process of a biometric authentication system based on FHE, highlighting potential adversarial actions. $A$ refers to the template reconstruction attack, and $B$ refers to the facial image reconstruction attack.}
        \label{fig:enrollment}
    \end{subfigure}
    
    \vspace{1em}  % A little vertical space to separate the subfigures
    
    % Subfigure (b) - Authentication process
    \begin{subfigure}[b]{0.5\textwidth}
        
        \includegraphics[width=\textwidth]{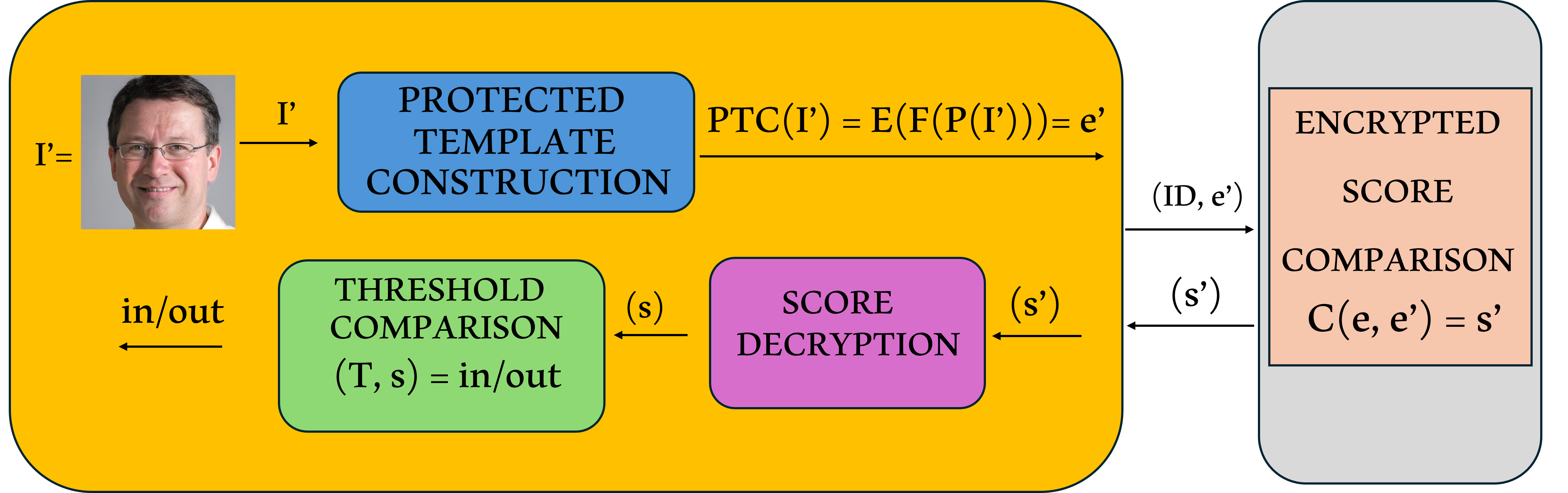}
        \caption{Traditional authentication process of a biometric system based on FHE.}
        \label{fig:authentication1}
    \end{subfigure}

    \vspace{1em}  % A little vertical space to separate the subfigures
    
    % Subfigure (b) - Authentication process
    \begin{subfigure}[b]{0.5\textwidth}
        
        \includegraphics[width=\textwidth]{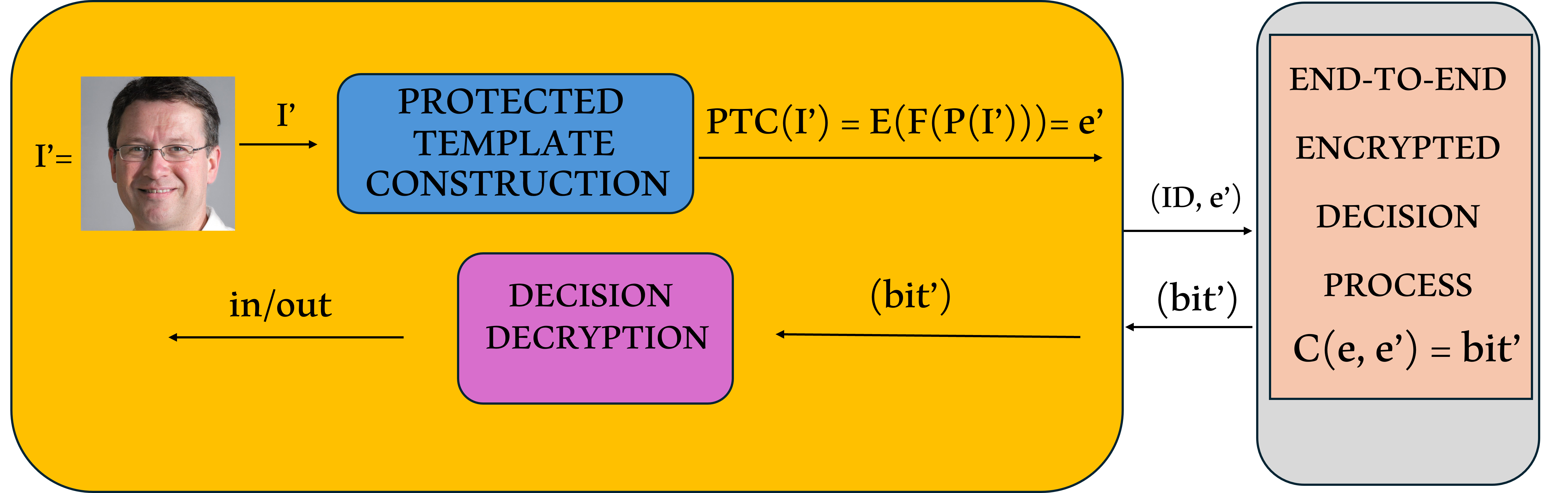}
        \caption{Authentication based on FHE with end-to-end encrypted decision process.}
        \label{fig:authentication2}
    \end{subfigure}

    \caption{Overview of Biometric Authentication Systems Based on FHE.}
    \label{fig:overall}
\end{figure}

Consider for example a biometric authentication system that use FHE (e.g.,~\cite{bauspiess2022improved,boddeti2018secure, drozdowski2019application,pradel2021privacy}) as depicted in Figure~\ref{fig:overall}. A typical enrollment process in these systems (shown in Figure~\ref{fig:enrollment}) involves three steps: preprocessing, feature extraction, and protection. The system first takes a raw biometric sample (such as a facial image or fingerprint) and processes it. The processed sample is then used to create a biometric template (feature vector) by extracting features from the sample. This template is then protected by an FHE-encryption. The encrypted template, along with the identity, is stored in a database for future authentication. 

During authentication, a new sample is taken from the user and processed to create a protected template, similarly to the enrollment process. This protected template is sent along with the claimed identity to the matching module which stores the encrypted template. 

If the decryption key is also held by the matching module, then breaking into it, would allow decrypting the protected template~\cite{shahreza2024breaking}. Hence, we assume that the decryption key is with the sensor that took the new sample. This means that breaking into the matching module does not reveal the templates to the adversary.

The first generation of FHE-based template protection schemes returned the encrypted matching scores\footnote{Typically the squared Euclidean distance or cosine similarity.} between the two samples --- the stored template and the new one~\cite{bauspiess2022improved,boddeti2018secure,drozdowski2019application,engelsma2022hers,tamiya2021improved}. With advances in FHE technology and by better adapting the feature comparison to FHE, it was possible to design newer systems that only return the binary authentication answer. Specifically, in the newer systems~\cite{bassit2021fast,bassit2023improved,pradel2021privacy}, the reply is an FHE-encrypted binary reply (``yes'' or ``no''), i.e.,
whether the new and the previous samples are close enough. Both methods are shown in Figures~\ref{fig:authentication1} and~\ref{fig:authentication2}, respectively.

%Consequently, its adoption has grown over time and is expected to further increase due to its potential to provide an efficient solution that prevents the exposure of similarity scores during authentication.

In this work, we investigate the basic level of security that can be offered by a  biometric authentication systems that returns the minimal amount of information that can be returned, namely, whether the authentication query succeeded or failed. We note that 
 previous work in this line assumed that the adversary obtains access to the similarity score, which allows an hill-climbing attack~\cite{alberto2015privacy,maiorana2014hill, mandal2016comprehensive,sarkar2020review,simoens2012framework} or a non-adaptive approach~\cite{bassit2023template} that approximates the target by solving a constrained optimization problem formulated over random fake templates with their corresponding scores.  

We start by recalling how one can reconstruct templates with minimal reconstruction loss by solving the relevant system of equations (first presented by Mohanty in~\cite{mohanty2007scores}). We then show how can one extend this approach to deal with binary answers, i.e., the adversary can only observe whether an authentication attempt succeeded or failed (e.g., by observing whether the system gives access). As part of the verification of the quality of reconstruction, we integrate our attack with previous work~\cite{vendrow2021realistic} to offer an attack pipeline capable of reconstructing high-resolution facial images that successfully pass the system more than 98\% of the time.

%Additionally, we include an attack on FHE systems that uses leaked comparison scores. This attack combines the algebraic approach from~\cite{mohanty2007scores}, with the modern generative deep network~\cite{vendrow2021realistic} that is able of reconstructing high-resolution facial images from recovered templates. While this approach is similar to~\cite{bassit2023template} in some aspects, it achieves perfect success rate, improving over their results for the case of  $d+1$ authentication attempts for a $d$-dimension template.
\paragraph{Our Main Contributions:}
\begin{compactitem}
      \item The first attack capable of reconstructing protected biometric templates with negligible reconstruction loss using only binary authentication outcomes.
    %\item A scores to high-resolution facial image attack pipeline targeting both score-releasing and binary-output protected biometric systems. %with over 98\% success.
    \item We conduct extensive evaluation and experiments on protected systems, demonstrating that encryption-based protection alone is insufficient to guarantee the irreversibility of biometric templates.%, challenging common assumptions in the field.
\end{compactitem}
We note that the experiments of this work focused on FHE-based systems and we expect similar results to apply to other cryptographic protection mechanisms, e.g., the use of attribute-based encryption or secure multiparty computation.

\paragraph{Organization of the paper:} 
Section~\ref{sec:related_works} discusses the related work. Section~\ref{sec:Out_method} introduces our attack methodology in detail. 
Section~\ref{sec:Evaluation__metrics} describes the evaluation methods. The experimental verification of our attacks is presented in Section~\ref{sec:Evaluation}. Finally, Section~\ref{sec:Conclusion} concludes the paper and  suggests directions for future work. 

\section{Related Works}
\label{sec:related_works}

We focus on existing reconstruction attacks most relevant to our work, specifically score-based attacks. As we examplify our work on FHE-protected templates, we also discuss attacks specific to FHE systems.

Soutar~\cite{soutar2002biometric} was the first to propose an iterative template
adaptation scheme, popularly known as the hill climbing attack, to break into a biometric system based on match scores.
Hill-climbing attacks are performed by iteratively submitting synthetic representations of the attacked user’s biometrics until successful recognition is achieved. At each step, the data is modified based on the results of the previous attempts, expressed in terms of matching scores known to the attacker, with the aim of improving the matching output.

Mohanty et al.~\cite{mohanty2007scores}  proposed a non-adaptive method that exploits the exposed similarity scores to construct equations from the revealed data. The template was recovered by solving the system of equations. The inverse of the affine transformation model was used to reconstruct facial images, leveraging released scores to simulate the behavior of the face recognition system.

The above methods were developed even before deep learning made a huge change in the field of biometrics, but their basic idea (hill climbing and solving a set of equations) is still applicable for systems that release scores. This include also systems that were protected using FHE. We now quickly recall specific attacks which are specific for FHE-protected systems that return scores (such as the first generation systems of~\cite{boddeti2018secure, bauspiess2022improved,drozdowski2019application,morampudi2021secure,yang2020secure}, all of which return scores).

Recently,~\cite{bassit2023template} proposed a non-adaptive attack on recovering raw templates from leaked comparison scores in systems that use holomorphic encryption (HE) for template protection. It focused on cosine similarity scores leaked from HE-based biometric systems,
which involve quantization
techniques to adapt the feature values to the HE plaintext
space. We note that our experiments included both cosine similarity and Euclidean distance scores, covering a wider range of systems; and the more advanced CKKS scheme~\cite{cheon2017homomorphic} for HE that supports floating-point encryption, and does not need quantization. The approach in~\cite{bassit2023template} used a random set of fake templates and their matching scores with the target template to construct a constrained optimization problem to recover the target template (with norm~1). The constrained optimization can include a varying number of fake templates, even less than the template's dimension. In this case the approximation of the template is less accurate. It achieves the perfect attack success rate (with no quantization) when the number of fake templates almost doubles the dimension of the template.  

Shahreza and Marcel~\cite{shahreza2024breaking} proposed a method for reconstructing face images from protected facial templates, including those protected by homomorphic encryption. Their attack assumes that the adversary has a complete knowledge of the protection scheme and its secrets, i.e., the decryption key. Thus, the adversary can decrypt the templates and analyze the unprotected templates. 

Recent works (which we call the second generation FHE-solutions) inmproved the running times of the FHE-computation, and were able to produce FHE-protected systems that only return the binary result of the authentication, such as the works of~\cite{bassit2021fast,bassit2023improved,pradel2021privacy}. This raises the question whether one can still reconstruct the templates when the adversary is given only access to whether the authentication query was successful or not.

% While sending a boolean result to the client is evidently more secure than transmitting a similarity score, as suggested in previous works~\cite{alberto2015privacy}, it is also argued that the vulnerability from exposing the similarity score in these systems is limited, though it can still be exploited for reverse attacks, such as hill-climbing attacks~\cite{alberto2015privacy,maiorana2014hill,sarkar2020review,mandal2016comprehensive,simoens2012framework}.

\section{Attack Methodology}
\label{sec:Out_method}
After defining the Adversarial model in Section~\ref{sec:sub:model}, we present a method for recovering biometric templates from similarity scores (Section~\ref{sec:Out_method_1}). While variants of score-based attacks have been studied in a prior work~\cite{bassit2023template}, our version offers improvements in certain settings and employs a unified algebraic framework. Importantly, a key component of this method is later adapted in our primary contribution: an advanced attack that reconstructs biometric templates using only binary (yes/no) matching outcomes (Section~\ref{sec:yes_no_attack}). Including the score-based attack serves both to provide completeness and to lay the groundwork for the more challenging binary outcome scenario. Section ~\ref{sec:end_to_end} presents the attack pipeline for facial image reconstruction.

\subsection{Adversarial Model}
\label{sec:sub:model}

We consider a threat model in which the feature extractor is run by a semi-honest service provider—assumed to follow the protocol correctly but potentially curious about inputs and outputs. The service provider holds the original templates which are encrypted under the matching module's (or sensor's) public-key, and cannot decrypt those templates. The attacker has a black-box access to the matching module and can submit arbitrary feature vectors, either derived from actual biometric samples or crafted independently. This is a common adversarial model (used in~\cite{bassit2023template,boddeti2018secure,drozdowski2019application,engelsma2022hers}), which assumes the adversary controls the sensor or has access to the feature extractor model in it.  The attacker has a limited number of queries, and depending on the attack scenario receives either similarity scores or binary match/no-match outcomes from the system.

The attack in Section~\ref{sec:Out_method_1} targets the score-releasing authentication system shown in Figure~\ref{fig:authentication1}. The attack in Section~\ref{sec:yes_no_attack} targets the binary-only system depicted in Figure~\ref{fig:authentication2}, where no data is revealed during the authentication process beyond whether the attempt was successful. We assume the system's false match rate (FMR) can be reasonably approximated, allowing the attacker to infer the underlying decision threshold.

In addition, both attacks rely on the following assumptions:
\begin{compactitem}
    \item The target template is stored protected on the server and cannot be accessed directly.
    \item The attacker requires black-box access to the feature extractor, in order to extract features from candidate images and to reconstruct images corresponding to recovered templates.
\end{compactitem}

\subsection {Reconstructing Protected Biometric Templates From Scores}

\label{sec:Out_method_1}

Let $f = [f^1, f^2, \ldots, f^d] \in \mathbb{R}^d$ represent the target user's biometric template (feature vector) created during the enrollment process without any protection and subsequently stored in a protected form in the database on the server side, where $d$ is the feature vector dimension. We aim to find an approximation to $f$.
Let $\{q_i\}_{i=1}^{d+1}$%$Q_1, Q_2, \ldots, Q_{d+1}$ 
be $d + 1$ independent feature vectors, where $q_i = [q_i^1, q_i^2, \ldots, q_i^d] \in \mathbb{R}^d$ 

% and $I_{i}$ is the corresponding biometric sample (an attacker can choose any biometric sample $I_{i}$ and save its features $q_i$ by extracting them from the sample).

\noindent\textit{Squared Euclidean distance case}: When the target system’s similarity score is the squared Euclidean distance, we follow the attack in~\cite{mohanty2007scores} based on the triangulation principle.  
For each $q_i$ we submit an authentication trial with the claimed identity being the target user's ID, totaling ${d + 1}$ independent trials without specific inputs (fewer attempts can be made at the expense of reduced accuracy). Let %$D_1,\ldots, D_{d+1}$ 
$\{s_i\}_{i=1}^{d+1}$ represent the squared Euclidean distances between $q_i$ and $f$, exposed during the authentication trials. %, where  $D_i$ is the squared Euclidean distance between $Q_i$ to $F$. 
Using this data, we can construct the following nonlinear system of equations: 
\begin{equation}
\label{eq:system}
\small
(f^1 - q_{i}^1)^2 +  \ldots + (f^d - q_{i}^d)^2 = s_i \quad i = 1, \dots, d+1
\end{equation}
%\begin{equation}
%\label{eq:system}
%\small
%\begin{gathered}
%(F_1 - Q_{11})^2 + (F_2 - Q_{12})^2 + \ldots + (F_d - Q_{1d})^2 = D_1 \\
%(F_1 - Q_{21})^2 + (F_2 - Q_{22})^2 + \ldots + (F_d - Q_{2d})^2 = D_2 \\
%\vdots \\
%(F_1 - Q_{(d+1)1})^2 + (F_2 - Q_{(d+1)2})^2 + \ldots + (F_d - Q_{(d+1)d})^2 = D_{d+1}
%\end{gathered}
%\end{equation}
which can be rewritten as:

\begin{equation}
\label{eq:system2}
\|f\|^2 - l_i = 0, \quad i = 1, \dots, d+1
\end{equation}
%
%\begin{equation}
%\small
%\label{eq:system2}
%\begin{gathered}
%\sum_{j=1}^d (F_j)^2  -L_1 = 0 \\
%\sum_{j=1}^d (F_j)^2 -L_2 = 0 \\
%\vdots \\
%\sum_{j=1}^d (F_j)^2 -L_{d + 1} = 0
%\end{gathered}
%\end{equation}
where $\|f\|^2=\sum_{j=1}^d (f^j)^2 $ and $l_i = s_i -\sum_{j=1}^d -2f^j q_{i}^j + (q_{i}^j)^2$ is the linear part of equation $i$.

\noindent Subtracting one of these equations, e.g., the last one, from all the others creates a linear system of equations of $d$ equations and $d$ variables as follows.
\begin{equation}
\label{eq:system3}
l_{d+1} - l_i = 0, \quad i = 1, \dots, d
\end{equation}
%\begin{equation}
%\label{eq:system3}
%\small
%\begin{gathered}
%L_{d+1} - L_1 = 0 \\
%L_{d+1} - L_2 = 0 \\
%\vdots \\
%L_{d+1} - L_d = 0
%\end{gathered}
%\end{equation}
%The linear equation system (Equation \ref{eq:system3}) created by subtracting the last equation in the non-linear equation system (Equation \ref{eq:system2}) from all the other equations. 
% It is equivalent to solving for 
% $F$ when:
% \begin{equation}
% \label{eq:system3}
% \begin{gathered}
% AF = B
% \end{gathered}
% \end{equation}
% given $A \in \mathbb{R}^{d \times d}$ and $B \in \mathbb{R}^d$.

An approximation for $f$ can be efficiently obtained by solving the linear equation system (Eq.~\eqref{eq:system3}).%, e.g., by using the least square algorithm~\cite{jiang1998least}. {\bf Orr: I must admit that I don't see why we bother in telling how to solve a system of linear equations. A reader of this paper who never heard that one can solve linear equations won't understand a thing anyway, and some reviewer may ask why Jiang's work is the one we picked of all methods.}

\noindent\textit{Cosine similarity case}: In this case, every authentication trial automatically creates a linear equation derived from the cosine similarity definition (Eq.~\eqref{eq:cosine_similarity_def}) and the fact that the deep feature vectors are normalized, so their magnitudes are $1$. 
\begin{equation}
\small
\label{eq:cosine_similarity_def}
\begin{gathered}
\text{cosine similarity} = \frac{A \cdot B}{\|A\| \|B\|}
\end{gathered}
\end{equation}
In this case, an attacker can use only $d$ authentication trials to creates a system of linear equations composed of $d$ equations and $d$ variables. The equation $i$ can be written as:
\begin{equation}
\label{eq:cosine_similarity}
\small
\begin{gathered}
c_{i} = {\sum_{j=1}^{d} f^{j} q_{i}^j}
\end{gathered}
\end{equation}
where $c_i$ is the cosine similarity score between $f$ and $q_i$.

We note that this approach offers a more accurate reconstruction than the one used in~\cite{bassit2023template}. Once enough equations are available, the solution is immediate (and accurate), unlike the approach of~\cite{bassit2023template} that may require additional information for a solution.

\subsection{Reconstructing Protected Biometric Templates
Using Only Binary Authentication Results}
\label{sec:yes_no_attack}

Let $T$ denote that system's matching threshold. We observe that the problem of recovering the target template using yes/no answers is equivalent to the following geometric problem: finding the center of a $d$-dimensional sphere with radius $T$ and a function \( A: \mathbb{R}^d \to \{0,1\} \) that maps a $d$-dimensional real-valued input \( x \in \mathbb{R}^d \) to either 0 or 1, i.e.,
\begin{equation}
\label{auth_functions}
A(x) =
\begin{cases}
    1, & \text{if } x \text{ is inside the sphere}, \\
    0, & \text{otherwise}.
\end{cases}
\end{equation}

\textit{Baseline approach:} Perform authentication attempts using a ``breaking-set'' that contains different facial images, and average all false match feature vectors that were successfully authenticated, as shown in Algorithm~\ref{alg:naive}. 

\begin{algorithm}
\caption{Baseline Approach }\label{alg:naive}
\begin{algorithmic}[1]
    \REQUIRE  function $A$ as defined in Eq.~\eqref{auth_functions} and a ``breaking-set'' $B$.
    \STATE $S = \{\}$
    \FOR{$b$ in $B$}
            \IF{$A(b)$}
                \STATE Set $S = S \cup \{b\}$
        \ENDIF
    \ENDFOR
    \STATE Return average of all facial templates in $S$ 
\end{algorithmic}
\end{algorithm}

% \begin{figure*}[ht]
%     \centering
%     \includegraphics[width=\textwidth]{yes_no_attack.png}
%     \captionsetup{justification=centering}
%     \caption{}
%     \label{fig:yes_no_attack}
% \end{figure*}

\begin{figure*}[ht]
    \centering
    \includegraphics[width=0.9\textwidth]{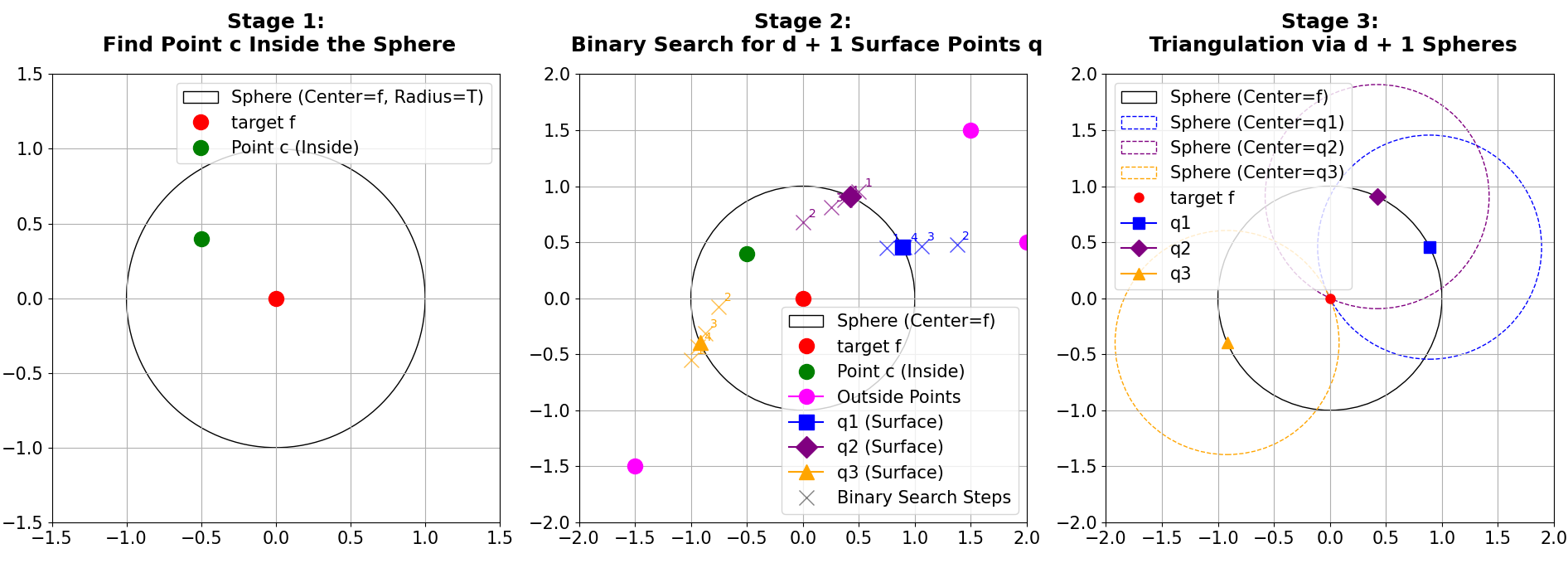}
    \captionsetup{justification=centering}
    \caption{Illustration of our template reconstruction attack in using only binary authentication results in 2D.}
    \label{fig:yes_no_attack}
\end{figure*}

\textit{Our Method:}
We suggest a superior method described in Algorithm~\ref{alg:yes_no} and illustrated in Figure~\ref{fig:yes_no_attack}.
The first stage of the attack is to find one feature vector $c$ that passes authentication. On average, this stage requires $1/\mathrm{FMR}$ authentication attempts (as defined by the False Match Rate).

The second stage focuses on finding $d + 1$ points on the surface of the sphere. We find a point $o_i$ outside the sphere by constructing a vector of length $2T$ in a random direction and adding it to $c$, then preform a binary search using the binary outputs of the system (function $A$ as defined in Eq.~\eqref{auth_functions}) for $P$ iterations using the binary results of the biometric authentication system.  We can see that as $P$ increases, the point $q_i$ (the binary search result) approaches the surface of the sphere, meaning that the distance between $q_i$ and $f$ converges to\footnote{We note that the distance between $q_i$ and $f$ after $P$ iterations of the binary search is in the range $[T-T/2^P,T+T/2^P]$.} to $T$. Repeating this process $d + 1$ times allows us to construct the non-linear system of equations given in Eq.~\eqref{eq:system}, where $s_1, s_2, \ldots, s_{d+1}$ are all equal to $T$. At the final stage, we solve the system of equations, as demonstrated in Section~\ref{sec:Out_method_1}. It is worth noting that the attacker does not need to know the value of $T$ for this stage because $T$ is eliminated during the algebraic solution by subtracting one of the equations from all the others.

\begin{algorithm}
\caption{Our Method}\label{alg:yes_no}
\begin{algorithmic}[1]
    \REQUIRE Precision parameter $P$, function $A$ as defined in Eq.~\eqref{auth_functions}, radius $T$, a ``breaking-set'' $B$. 
    \FOR{$b$ in $B$}
        \IF{$A(b)$} 
            \STATE $c = b$; break;
        \ENDIF
    \ENDFOR
    \STATE Initialize an empty system of equations $S$ (like in Eq.~\eqref{eq:system})\!\!\!
        \FOR{$d + 1$ iterations}
            \STATE Sample a random vector $r$ with norm $|r|=1$
            \STATE  $o_i$ = $c + r(2T)$
            \STATE $c_i = c$
            \FOR{$P$ iterations}
            \STATE $q_i = (c_i + o_i) / 2$
            \IF{$A(q_i)$}
                \STATE $c_i = q_i$
            \ELSE
                \STATE $o_i = q_i$
            \ENDIF
        \ENDFOR
    \STATE Add $(c_i+o_i)/2$ to $S$ with distance $T$ from center
    \ENDFOR
    \STATE Solve $S$ for the center of the sphere
    \STATE \textbf{return} the solution of $S$
\end{algorithmic}
\end{algorithm}

The expected number of authentication attempts needed for reconstruction with a negligible reconstruction loss is $\frac{1}{\text{FMR}} \ + P \cdot (d + 1)$.

\subsection{Facial Image Reconstruction Attack Pipeline}
\label{sec:end_to_end}
The goal of the attack is to extract a close approximation to the biometric sample (facial image) that was sampled during
the enrollment process of a protected biometric authentication system. We designed a scores to high-resolution facial image attack pipeline by using the approach for reconstructing a close approximation $\hat{f}$ of the original facial template $f$ followed by a image-from-template reconstruction method proposed in~\cite{vendrow2021realistic}.\footnote{Realistic Face Reconstruction implementation \url{https://github.com/evendrow/face-reconstruction}.} The method in~\cite{vendrow2021realistic} proposes an efficient search algorithm that aims to minimize the distance between $F(G(x))$ and a target template $t$, where $x$ is a latent vector, $G$ is a high-resolution face image generator based on a Generative Adversarial Network (GAN), and $F$ is a feature extractor. The objective is:
\[
x^* = \arg\min_x \; \mathcal{D}(F(G(x)), t), \quad \hat{x} = G(x^*),
\]
where $\mathcal{D}$ is a distance metric such as cosine similarity or squared Euclidean distance. The use of a GAN enables reconstruction of high-resolution facial images. Given a reconstructed template from the first part of the attack and assuming black-box access to the feature extractor, the attacker can directly apply this method without requiring any additional training.

\section{Evaluation metrics}
\label{sec:Evaluation__metrics}
Reversing attacks against protected biometric authentication systems are challenging and depend on various factors that influence their practicality. To facilitate effective comparison across different attack types and to comprehensively evaluate the security of such systems, we propose a framework that assesses both the feasibility and quality of attacks. This framework considers key attributes affecting attack viability and includes empirical metrics to evaluate the quality of the reconstructed outputs.

\textbf{Template Reconstruction Quality}
\begin{compactitem}
   \item Reconstruction Loss: Reconstruction loss is the similarity score between the reconstructed template and the original template. The closer the reconstruction loss is to 0, the higher the likelihood of performing a successful authentication based on the reconstructed template. Additionally, a reverse attack from the biometric template to the biometric sample is more likely to succeed as it will be based on a vector which is closer to the ground truth (the template of the real user).

   \item Number of Authentication Trials: The number of authentication trials required for an attack should be as low as possible. The higher the number of authentication trials, the lower the likelihood of a successful attack due to defense mechanisms that limit the number of trials or detect unusual authentication attempts~\cite{simoens2012framework}.

  \item Time: The attack time should be practical and as low as possible. When comparing the time between different attacks, it is important to have equal conditions, such as the same hardware and targeting the same system.
\end{compactitem}
\textbf{Facial Image Reconstruction Quality}

%{\bf Orr: proposal for a more concise write-up:}
We adopt the 4 scenarios described in~\cite{gomez2020reversing} for the evaluation of reversing attacks to the settings of reconstruction facial images. The reconstructed image is tested in a different system. This new system may either have used the same acquisition image (e.g., when the user supplies the image), or a different one (e.g., when a new image is taken). The new system may also use the same feature extractor or a different one. Each of the four scenarios are thus: Same Image-Same Feature Extraction (SISFE, corresponding to scenario~1 of~\cite{gomez2020reversing}), Different Image-Same Feature Extraction (DISFE, corresponding to scenario~2 of~\cite{gomez2020reversing}), Same Image-Different Feature Extraction (SIDFE, corresponding to scenario~3 of~\cite{gomez2020reversing}), Different Image-Different Feature Extraction (DIDFE, corresponding to scenario~3 of~\cite{gomez2020reversing}).

\section{Experiments and Results}
\label{sec:Evaluation}

In this section, we present our comprehensive evaluation. To exemplify the strength of our method, we apply it to templates that were protected using FHE, as these are the most protected systems that we have available today.
Section~\ref{sec:Evaluation_1} evaluates the Template Reconstruction Attacks from Scores. 
Section~\ref{sec:Evaluation_2}
focuses on evaluating the Template Reconstruction Attacks from Binary Results. 
Section~\ref{sec:Evaluation_3}
focuses on evaluating the facial images reconstructed from the reconstructed facial templates (the full attack pipeline).

\textit{Users}: The LFW~\cite{huang2008labeled} data set includes
13,233 face images of 5,749 identities and it is one of
the most popular benchmarks for face verification. Out
of the 1,680 identities which have two or more images,
we randomly picked 300 identities as the users of the system. 
% The ``breaking-set'' consists of all facial images of users who are not in the target set.

We used a machine with an Intel i7-8700 processor running at 3.2 GHz.
\subsection{Evaluation of the Templates Reconstruction Attacks from Scores}
\label{sec:Evaluation_1}
\textit{Target Systems}: For the case of cosine similarity our target system is the well-known open source biometric authentication ``Secure Face Matching'\footnote{Secure Face Matching \url{https://github.com/human-analysis/secure-face-matching}.}~\cite{boddeti2018secure} that provides enrollment and authentication functionality. For the case of squared Euclidean distance we follow the approach of previous works~\cite{drozdowski2019application, bauspiess2022improved}. These works use the CKKS fully homomorphic encryption scheme~\cite{cheon2017homomorphic} implemented by TenSEAL\footnote{TenSEAL  implementation \url{https://github.com/OpenMined/TenSEAL}.}~\cite{benaissa2021tenseal} for template protection. For feature extraction, we used standard algorithms and implementations, specifically ArcFace\footnote{{ArcFace implementation \url{https://github.com/peteryuX/arcface-tf2}.}}~\cite{deng2019arcface} and FaceNet\footnote{FaceNet implementation \url{https://github.com/timesler/facenet-pytorch}.} that are also used in~\cite{boddeti2018secure, drozdowski2019application, bauspiess2022improved}.
We determined the system thresholds for False Match Rates (FMRs) of 1.0\% and 0.1\% using a standard method and used the standard facial image dataset LFW~\cite{huang2008labeled} as used in~\cite{boddeti2018secure,jindal2020secure, ma2017secure}.
 
We reconstructed facial image templates of enrolled users using the algebraic approach described in Section \ref{sec:Out_method_1} (where $d = 512$ represents the feature vector dimension of ArcFace and FaceNet) and using a hill-climbing attack.

\textit{Implementation Details}: For the algebraic attack, we conducted authentication trials using the random data and then follow the process discribded in Section \ref{sec:Out_method_1}. 

For the hill-climbing attack, we followed the approach outlined in previous works~\cite{maiorana2014hill,alberto2015privacy}. We iteratively submitted synthetic templates of the attacked user’s biometrics, modifying the data at each step based on the results of the previous attempts. The data modification was controlled by the step-size hyperparameter, which we set to $0.07$ after optimizing its value.

We performed the attack for up to 4,000 authentication trials, which is approximately eight times the number of trials used in our attack to achieve the same goal.

\begin{table}[ht]
    \caption{Results of algebraic and hill-climbing approaches against the target FHE-protected biometric authentication system that releases squared Euclidean distance matching scores.}
    \Large
    \resizebox{\columnwidth}{!}{%
    \begin{tabular}{lcccccccc}
        \toprule
        \textbf{Attack Type} & \multicolumn{4}{c}{\textbf{Algebraic Approach}} & \multicolumn{4}{c}{\textbf{Hill-Climbing}} \\
        \cmidrule(lr){2-5} \cmidrule(lr){6-9}
        {\textbf{Feature Extractor}} & \multicolumn{2}{c}{\textbf{ArcFace}} & \multicolumn{2}{c}{\textbf{FaceNet}} & \multicolumn{2}{c}{\textbf{ArcFace}} & \multicolumn{2}{c}{\textbf{FaceNet}} \\
        \midrule
        \textbf{Auth. Attempts} & \multicolumn{4}{c}{513} & \multicolumn{4}{c}{4000}\\
        \midrule
        \textbf{Time (sec)} &  \multicolumn{4}{c}{13} & \multicolumn{4}{c}{98} \\
        \midrule
        \textbf{Reconst. Loss} & \multicolumn{2}{c}{$1.7 \times 10^{-26}$} & \multicolumn{2}{c}{$8.5 \times 10^{-27}$} & \multicolumn{2}{c}{$0.332 \pm 0.05$} & \multicolumn{2}{c}{$0.32 \pm 0.04$} \\
        \bottomrule
    \end{tabular}%
    }
    \label{tab:table2}
\end{table}

\begin{table}[ht]
    \caption{Results of algebraic and hill-climbing approaches against the FHE-protected ``Secure Face Matching'' biometric authentication system that releases cosine similarity matching scores.}
    \Large
    \centering
    \resizebox{\columnwidth}{!}{%
    \begin{tabular}{lcccccccc}
        \toprule
        \textbf{Attack Type} & \multicolumn{4}{c}{\textbf{Algebraic Approach}} & \multicolumn{4}{c}{\textbf{Hill-Climbing Approach}} \\
        \cmidrule(lr){2-5} \cmidrule(lr){6-9}
        \textbf{Feature Extractor} & \multicolumn{2}{c}{\textbf{ArcFace}} & \multicolumn{2}{c}{\textbf{FaceNet}} & \multicolumn{2}{c}{\textbf{ArcFace}} & \multicolumn{2}{c}{\textbf{FaceNet}} \\
        \midrule
        \textbf{Auth. Attempts} & \multicolumn{4}{c}{512} & \multicolumn{4}{c}{4000} \\
        \midrule
        \textbf{Time (sec)} & \multicolumn{4}{c}{$465$} & \multicolumn{4}{c}{$3692$} \\
        \midrule
        \textbf{Reconst. Loss} & \multicolumn{2}{c}{$0.061 \pm 0.007$} & \multicolumn{2}{c}{$0.069 \pm 0.023$} & \multicolumn{2}{c}{$0.35 \pm 0.074$} & \multicolumn{2}{c}{$0.351 \pm 0.075$} \\
        \bottomrule
    \end{tabular}%
    }
    \label{tab:table_new}
\end{table}

 \textit{Results}:
To evaluate the results, we used the comprehensive evaluation method proposed in Section \ref{sec:Evaluation__metrics}. The reconstruction loss for the algebraic approach was negligible in both cases, less than $1.7 \times 10^{-26}$ and $0.07$ for the case of squared euclidean distance and cosine similarity, respectively (The difference between cosine similarity and SED results is due to lower FHE noise in the newer SED-based system). While the average reconstruction loss for the hill-climbing was less than $0.33$ and $0.351$ for the case of squared euclidean distance and cosine similarity.

When the threshold of the systems was set for an FMR of 1\% the thresholds were 1.444 and 1.243 for ArcFace and FaceNet, respectively. For an FMR of 0.1\% the thresholds were 1.18 and 0.967 for ArcFace and FaceNet, respectively. In the case of cosine similarity, for an FMR of 1\%, the thresholds were 0.411 and 0.377 for ArcFace and FaceNet, respectively. For an FMR of 0.1\%, the thresholds were 0.364 and 0.325 for ArcFace and FaceNet, respectively. The attack's results for the squared euclidean distance and cosine similarity are present in Table \ref{tab:table2} and \ref{tab:table_new}, receptively. It is worth noting that 100\% of the reconstructed templates achieve a reconstruction loss below the system threshold using only 513 authentication attempts. In comparison, the work of~\cite{bassit2023template} conducted a similar experiment on the LFW dataset with ArcFace features and cosine similarity, achieving the same success rate but requiring 700 attempts.

\subsection{Evaluation of the Templates Reconstruction Attack From Binary Results}
\label{sec:Evaluation_2}

\textit{Target Systems}: We utilized the feature extractors and thresholds described in Section \ref{sec:Evaluation_1}.  
Facial image templates of enrolled users were reconstructed using our method, as detailed in Section \ref{sec:yes_no_attack}, where $d = 512$ represents the feature vector dimension of ArcFace and FaceNet.  
For comparison, we employed the Baseline approach described in Section \ref{sec:yes_no_attack} as a baseline.

\textit{Implementation Details}: 
We set \( P = 20 \), so the number of authentication attempts required by our method is given by \( P \cdot (d + 1) = 20 \cdot 513 + \frac{1}{\text{FMR}} \) on average.  
For a fair comparison, the Baseline approach was allotted the same number of authentication attempts. In this case the ``breaking-set'' consisted of all images that did not belong to the target identity. The total number of attempts for both methods averaged 10,360 and 11,260 in the cases of 1\% and 0.1\% FMR, respectively. The attack runtime was under 2 seconds in both methods. This time reflects only the algorithmic computation time, as we did not perform the full end-to-end encrypted decision process.

\textit{Results}: 
The reconstruction loss for our method was negligible, less than  $7 \times 10^{-5}$. While the average reconstruction loss for the baseline approach was at least 0.3—ranging from 29\% and 39\% of the system’s threshold, this is more than $4,200$ times greater than our reconstruction loss using the same number of authentication attempts. The attack's results are presented in Table \ref{tab:yes_no_table}. 

\begin{figure}[ht]
    \centering
    \includegraphics[width=0.45\linewidth]{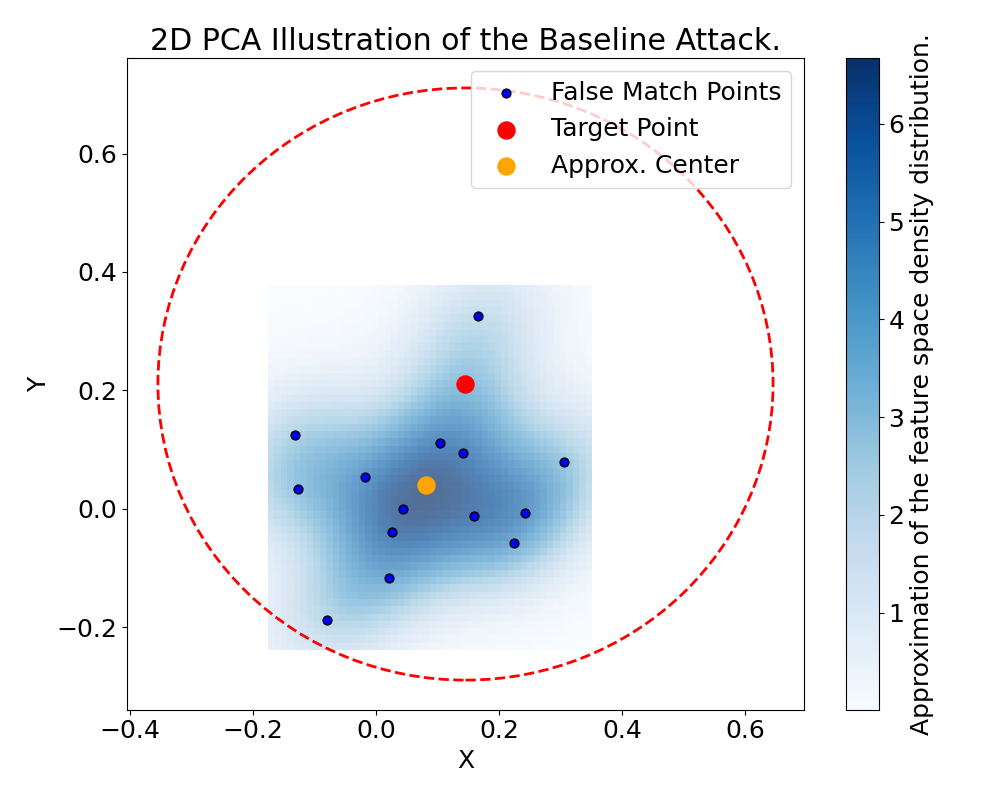}
     \includegraphics[width=0.45\linewidth]{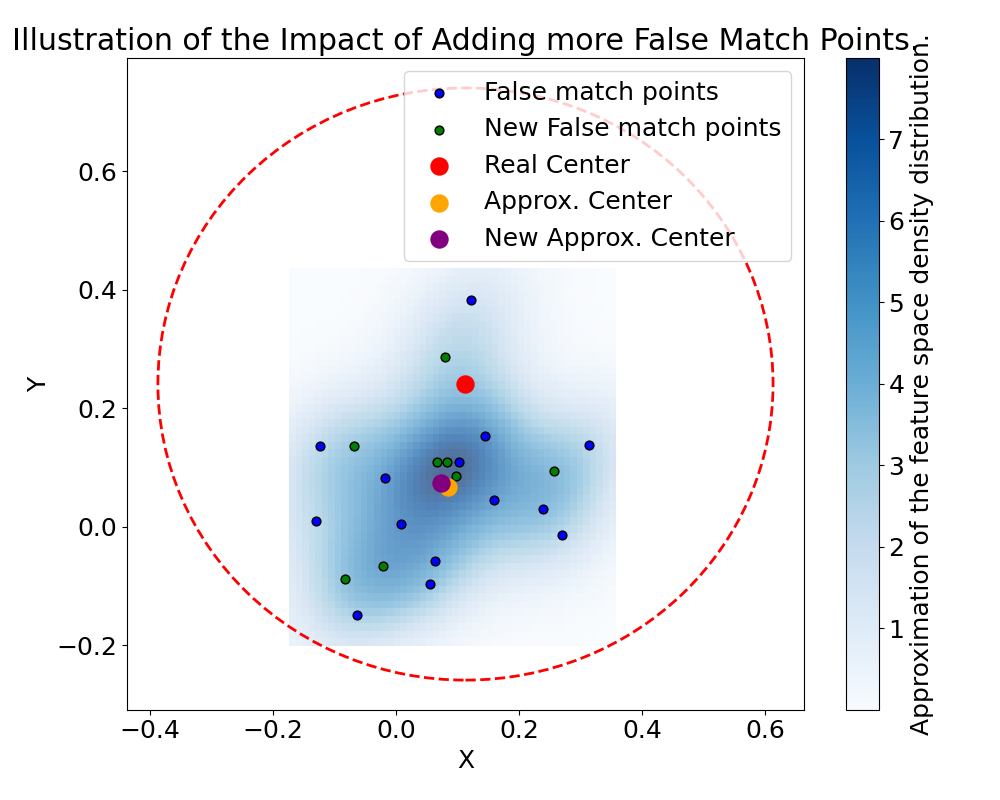}
    \caption{Illustration of the baseline attack using 2D PCA: left -- the approximation of the feature space density distribution within the match space of the target template; right -- Illustration of the impact of adding more false match points.}
    \label{fig:pca1}
\end{figure}

\begin{figure}[ht]
    \centering
    \begin{minipage}{0.48\linewidth}
        \centering
        \includegraphics[width=\linewidth]{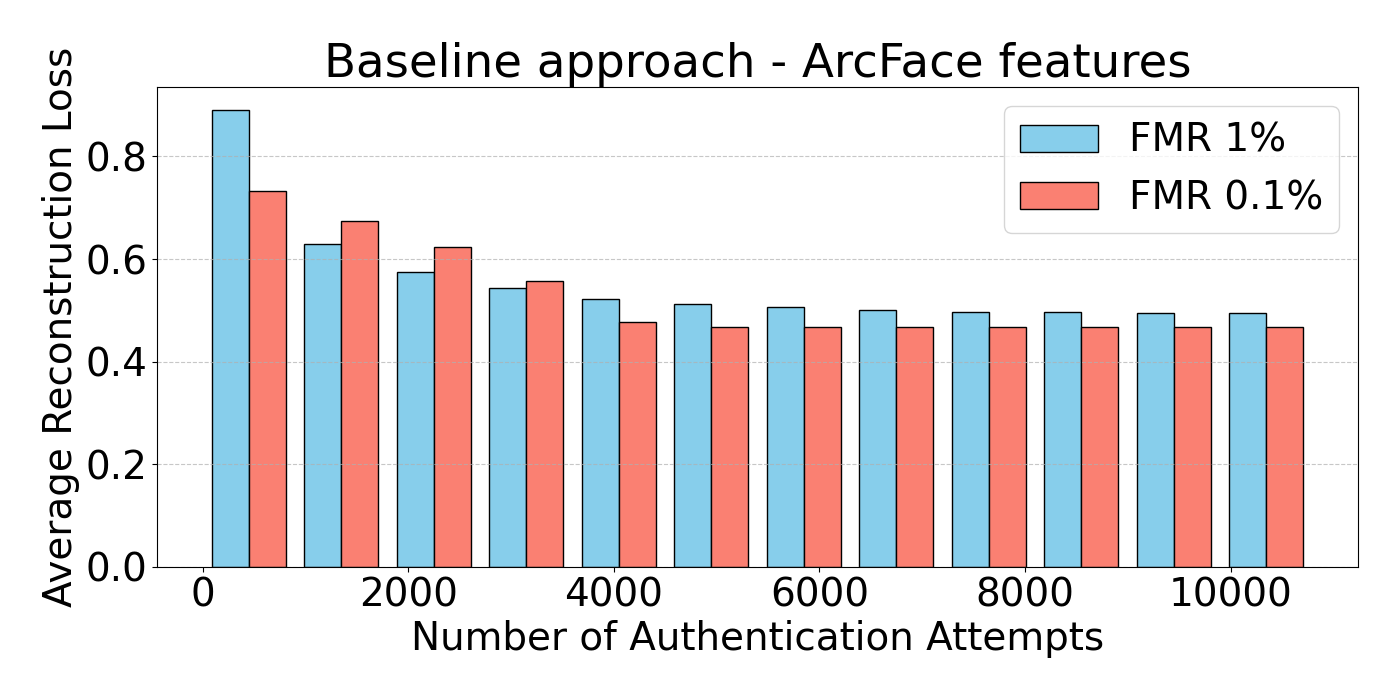}
        \caption{Reconstruction loss of ArcFace templates using the baseline approach.}
        \label{fig:arcface_baseline}
    \end{minipage}
    \hfill
    \begin{minipage}{0.48\linewidth}
        \centering
        \includegraphics[width=\linewidth]{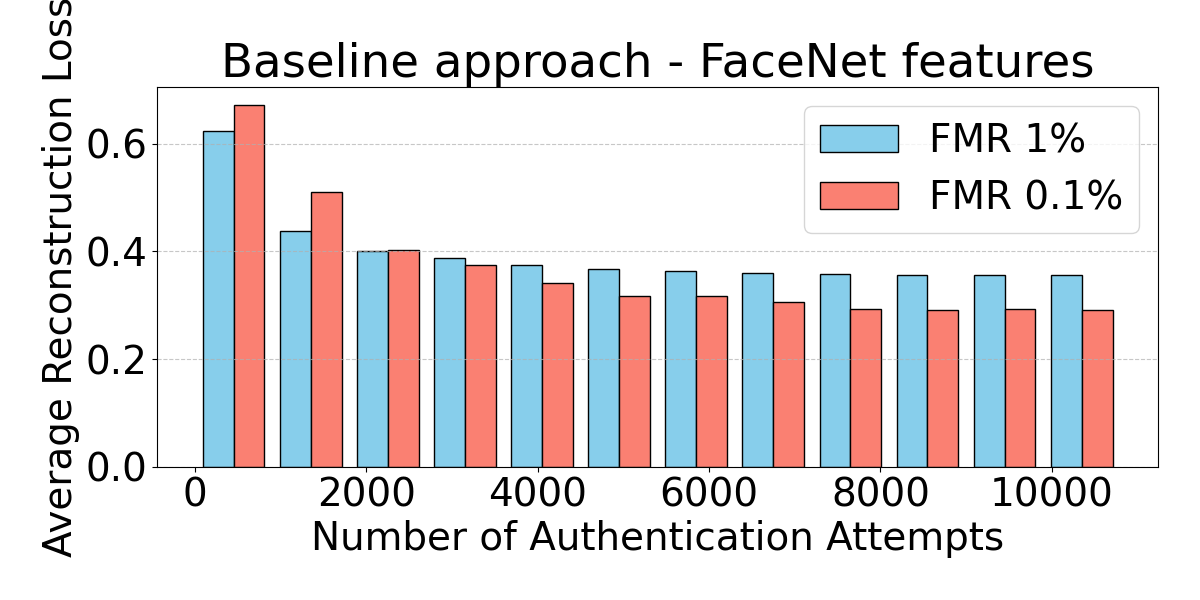}
        \caption{Reconstruction loss of FaceNet templates using the baseline approach.}
        \label{fig:facenet_baseline}
    \end{minipage}
\end{figure}

Our results demonstrate that our solution is the most efficient for achieving perfect reconstruction. Furthermore, we observed interesting insights regarding the baseline approach: while it may seem intuitive that averaging an increasing number of false match points would converge to the real template, this only holds if those points are uniformly distributed around the center of the sphere. However, this assumption does not hold in practice. The false match points are not only constrained to the sphere but are also distributed according to the underlying structure of the feature space. Figure \ref{fig:pca1}-left illustrates an approximation of the feature space density distribution within the match space of the target template in 2D using PCA. In Figure \ref{fig:pca1}-right, it can be seen that new false match points fall within the same feature space distribution, even though these new false match points are from different images of different identities. The improvement of the baseline approach over time can be seen in Figures \ref{fig:arcface_baseline} and \ref{fig:facenet_baseline}.

\begin{table}[ht]
    \caption{Results of our method compare to the Baseline approach of the reconstructed templates using only binary authentication results}
    \Large
    \centering
    \resizebox{\columnwidth}{!}{%
    \begin{tabular}{lcccccccc}
        \toprule
        \textbf{Attack Type} & \multicolumn{4}{c}{\textbf{Our Method}} & \multicolumn{4}{c}{\textbf{Baseline Approach}} \\
        \cmidrule(lr){2-5} \cmidrule(lr){6-9}
        \textbf{Feature Extractor} & \multicolumn{2}{c}{\textbf{ArcFace}} & \multicolumn{2}{c}{\textbf{FaceNet}} & \multicolumn{2}{c}{\textbf{ArcFace}} & \multicolumn{2}{c}{\textbf{FaceNet}} \\
        \midrule
        \textbf{FMR} & \textbf{1\%} & \textbf{0.1\%} & \textbf{1\%} & \textbf{0.1\%} & \textbf{1\%} & \textbf{0.1\%} & \textbf{1\%} & \textbf{0.1\%} \\
        \midrule
    
        \textbf{Reconst. Loss} & \multicolumn{2}{c}{$7 \times 10^{-5}$} & \multicolumn{2}{c}{$3.8 \times 10^{-5}$} & {$0.49\pm 0.02 $} & {$0.46 \pm 0.29  $} & {$0.35 \pm 0.02$} & {$0.30 \pm 0.16$} \\
    
        \bottomrule
    \end{tabular}%
    }
    \label{tab:yes_no_table}
\end{table}

\subsection{Evaluation of the Scores to Facial Image Attack}
\label{sec:Evaluation_3}

This section evaluates our complete attack from Section~\ref{sec:end_to_end}. Figure~\ref{fig:reconstruct} shows examples of reconstructed images compared to their corresponding target images. We tested the attack under the scenarios in Section~\ref{sec:Evaluation__metrics}. In the SIDFE/DIDFE experiments, the results for ArcFace correspond to reconstructions from protect ArcFace templates tested on FaceNet based system, while the results for FaceNet represent the opposite direction. Tables~\ref{tab:table4} and~\ref{tab:table5} report success rates against score-releasing systems based on squared Euclidean distance and the "Secure Face Matching" system (cosine similarity), respectively. Table~\ref{tab:table6} presents results for facial images generated from templates reconstructed via the attack in Section~\ref{sec:yes_no_attack}.

\begin{table}[ht]
    \centering
    \caption{Our Full Attack Pipeline Results against the target biometric authentication system based on FHE and Squared euclidean distance.}
    \label{tab:table4}
    \begin{tabular}{lccccc}
        \toprule
        \textbf{Attack Type/} & \multicolumn{2}{c}{\textbf{ArcFace}} & \multicolumn{2}{c}{\textbf{FaceNet}} \\
        \cmidrule(lr){2-3} \cmidrule(lr){4-5}
         \textbf{Target System} & \textbf{1\%} & \textbf{0.1\%} & \textbf{1\%} & \textbf{0.1\%} \\
        \midrule
        SISFE & 100\% & 99\%  & 100\% & 100\% \\
        DISFE & 97\% & 77.6\% & 99.3\% & 95.3\%  \\
        SIDFE & 85.3\% & 69.3\%  & 81\% & 50\% \\
        DIDFE & 80\% & 57\% & 67.6\% & 35.3\%  \\
        \bottomrule
    \end{tabular}
\end{table}

\begin{table}[ht]
    \centering
    \caption{Our Full Attack Pipeline Results against the ``Secure Face Matching'' biometric authentication system that
based on FHE and cosine similarity.}
    \label{tab:table5}
    \begin{tabular}{lccccc}
        \toprule
        \textbf{Attack Type /} & \multicolumn{2}{c}{\textbf{ArcFace}} & \multicolumn{2}{c}{\textbf{FaceNet}} \\
        \cmidrule(lr){2-3} \cmidrule(lr){4-5}
         {\bf Target System}& \textbf{1\%} & \textbf{0.1\%} & \textbf{1\%} & \textbf{0.1\%} \\
        \midrule
        SISFE & 99.6\% & 98.3\%  & 100\% & 100\% \\
        DISFE & 94.6\% & 73.6\% & 99\% & 94.6\%  \\
        SIDFE & 84.6\% & 66.6\%  & 79\% & 43.3\% \\
        DIDFE & 79\% & 52.6\% & 61.3\% & 24.6\%  \\
        \bottomrule
    \end{tabular}
\end{table}

\begin{table}[ht]
    \centering
    \caption{Our Full Attack Pipeline Results using only Binary Authentication Scores}
    \label{tab:table6}
    \begin{tabular}{lccccc}
        \toprule
        \textbf{Attack Type /} & \multicolumn{2}{c}{\textbf{ArcFace}} & \multicolumn{2}{c}{\textbf{FaceNet}} \\
        \cmidrule(lr){2-3} \cmidrule(lr){4-5}
         {\bf Target System}& \textbf{1\%} & \textbf{0.1\%} & \textbf{1\%} & \textbf{0.1\%} \\
        \midrule
        SISFE & 100\% & 99\%  & 100\% & 100\% \\
        DISFE & 96.3\% & 77.3\% & 99.3\% & 95\%  \\
        SIDFE & 85.3\% & 69\%  & 80.3\% & 48.6\% \\
        DIDFE & 80\%  & 55\% & 65\% & 33.6\%\\
        \bottomrule
    \end{tabular}
\end{table}

As can be seen, the reconstructed facial images successfully pass the system more than 98\%. Even in the hardest case of DIDFE, the success rate ranges from 80\% to 24.6\%. We remind the reader that this is 246 times larger than the FMR for this configuration.

\section{Conclusions and Future Work}
\label{sec:Conclusion}
This work reveals that biometric authentication systems, protected even with the most advanced cryptographic tools remain vulnerable, in configurations that release only binary match outcomes while the data itself remains protected. We demonstrate the first attack capable of reconstructing highly accurate biometric templates and corresponding realistic facial images under this limited feedback model, challenging the assumption that such systems offer strong irreversibility guarantees.

Our findings highlight the need to reassess the security of protected biometric systems, especially in scenarios where attackers can interact with the matching module post-feature-extraction.

One direction for future work is to improve the presented attacks, for example by analyzing the trade-off between reconstruction loss and the number of authentication attempts. Incorporating prior knowledge about the target distribution may reduce the number of required equations without degrading reconstruction quality -- a strategy shown effective in related tasks such as template-to-image inversion using deep generative models.

Further research should also consider the extension to other biometric modalities. Finally, it is important to evaluate systems that deploy combinations of defense mechanisms, to understand whether layered protections meaningfully mitigate the vulnerabilities identified in this work.

\section*{Acknowledgments}

The authors thank Adi Shamir for the technical discussions, Marc Stevens, Dennis Jackson, and David Niehues, for their suggestions. We also thank the anonymous reviewers for their insightful comments and suggestions.

{\small
\bibliographystyle{ieee}
\bibliography{egbib}
}

\end{document}